%% file: main.tex
\documentclass[conference]{IEEEtran}
\IEEEoverridecommandlockouts

\usepackage{cite}
\usepackage{amsmath, amssymb, amsfonts}
\usepackage{algorithmic}
\usepackage{graphicx}
\usepackage[many]{tcolorbox}
\usepackage{textcomp}
\usepackage{xcolor}
\usepackage{array}
\usepackage{rotating}
\usepackage{hyperref}
\usepackage{comment}
\usepackage{xspace}
\usepackage{listings}
\usepackage{longtable}
\usepackage{paralist}
\usepackage{blindtext}
\usepackage{wasysym}
\usepackage{pifont}
\usepackage{xurl}
\usepackage[para]{footmisc}
\usepackage{lscape}
\usepackage{threeparttable}
\usepackage{supertabular}
\usepackage{tablefootnote}

\def\BibTeX{{\rm B\kern-.05em{\sc i\kern-.025em b}\kern-.08em
    T\kern-.1667em\lower.7ex\hbox{E}\kern-.125emX}}

\input{commands}

\begin{document}

\title{On the Need for Configurable Travel Recommender Systems: A Systematic Mapping Study\\
}

\author{\IEEEauthorblockN{Rickson Simioni Pereira}
\IEEEauthorblockA{
\textit{GSSI, L'Aquila, Italy }\\
rickson.pereira@gssi.it}
\and
\IEEEauthorblockN{Claudio Di Sipio}
\IEEEauthorblockA{
\textit{University of l'Aquila, Italy}\\
claudio.disipio@univaq.it}
\and
\IEEEauthorblockN{Martina De Sanctis}
\IEEEauthorblockA{
\textit{GSSI, L'Aquila, Italy }
\\ martina.desanctis@gssi.it}

\and
\IEEEauthorblockN{Ludovico Iovino}
\IEEEauthorblockA{
\textit{GSSI, L'Aquila, Italy}\\
ludovico.iovino@gsss.it}
}

\maketitle

\begin{abstract}
Travel Recommender Systems (TRSs) have been proposed to ease the burden of choice in the travel domain, by providing valuable suggestions based on user preferences. Despite the broad similarities in functionalities and data provided by TRSs, these systems are significantly influenced by the diverse and heterogeneous contexts in which they operate. This plays a crucial role in determining the accuracy and appropriateness of the travel recommendations they deliver. For instance, in contexts like smart cities and natural parks, diverse runtime information—such as traffic conditions and trail status, respectively—should be utilized to ensure the delivery of pertinent recommendations, aligned with user preferences within the specific context. However, there is a trend to build TRSs from scratch for different contexts, rather than supporting developers with configuration approaches that promote reuse, minimize errors, and accelerate time-to-market. To illustrate this gap, in this paper, we conduct a systematic mapping study to examine the extent to which existing TRSs are configurable for different contexts. The conducted analysis reveals the lack of configuration support assisting TRSs providers in developing TRSs closely tied to their operational context. Our findings shed light on uncovered challenges in the domain, thus fostering future research focused on providing new methodologies enabling providers to handle TRSs configurations. 
\end{abstract}


\section{Introduction}
\label{sec:Introduction}
\input{src/Introduction}

\section{Motivation}
\label{sec:Motivation}
\input{src/Motivation}

\section{ Methodology and Results}
\label{sec:Methodology}
\input{src/Methodology}

\section{Challenges and Future Research Directions}
\label{sec:discussion}
\input{src/Discussion}

\section{Threats to validity}
\label{sec:Threats}
\input{src/Threats}

\section{Conclusion}
\label{sec:Conclusion}
\input{src/Conclusion}

\section*{Acknowledgments}

The work was partially supported by the research project RASTA: Realtà Aumentata e Story-Telling Automatizzato per la valorizzazione di Beni Culturali ed Itinerari; Italian MUR PON Proj. ARS01 00540, the MUR (Italy) Department of Excellence 2023 - 2027 for the GSSI, and the PRIN 2022 PNRR project \emph{``FRINGE: context-aware FaiRness engineerING in complex software systEms''} grant n. P2022553SL.

\bibliographystyle{IEEEtran}
\bibliography{bib.bib}

\end{document}

%% file: commands.tex
\newcommand*{\ie}{i.e.,\@\xspace}
\newcommand*{\eg}{e.g.,\@\xspace}

\newcommand*{\etal}{et al.\@\xspace}

\lstdefinestyle{searchstringstyle}{
  basicstyle=\ttfamily\scriptsize,
breaklines=true,                 
  captionpos=b,                    
  numbers=none,                    
  numbersep=5pt,                  
  showspaces=false,                
  showstringspaces=false,
  showtabs=false,                  
  tabsize=2,
  frame=single
}

\newtcolorbox{shadedbox}{
enhanced jigsaw,
colback=white,
boxrule=0.80pt,
left=0.3em,
right=0.3em,
top=0.1em,
bottom=0.05em
}

\newcommand{\rqfirst}{\textbf{RQ$_1$}: \textit{What are the algorithms used by the existing TRSs?}}
\newcommand{\rqsecond}{\textbf{RQ$_2$}: \textit{To which extent do existing TRSs handle real-time data to manage the dynamicity of their operational context?}}
\newcommand{\rqthird}{\textbf{RQ$_3$}: \textit{To which extent do the existing TRSs provide configuration features from the standpoint of systems providers?}}

%% file: src/Introduction.tex

Tourism activities have been remarkably affected by the evolution of modern information systems that adopt cutting-edge technologies, thus transforming traditional tourism systems into smart tourism ones~\cite{hamid_how_2021}. More importantly, travel activities strongly depend on the users' personal experiences~\cite{delic_research_nodate} and the type of personality~\cite{a9d77a28-8be9-31fc-842d-09259c364c5f}. 
In this respect, Travel Recommender Systems (TRSs)~\cite{ricci_travel_2002} are a special class of recommender systems \cite{Ricci2011} focused on providing relevant travel recommendations, \eg~suggesting Points of Interest (POIs) or touristic itineraries. A widely investigated aspect 
is the extraction of user preferences~\cite{Tkalcic2015} to recommend suitable items. 
%
Nevertheless, although considerable attention has been devoted to the end-users of TRSs and their preferences, we argue that the impact of the operational context of TRSs, and its dynamicity, on travel recommendations has been disregarded. Indeed, these systems are significantly influenced by the diverse and heterogeneous contexts in which they operate. 
For example, information that is crucial in one context—e.g.,~traffic conditions in a smart city—may be negligible in a different setting, like a natural park, where the state of trails 
would be more beneficial in providing pertinent recommendations. Moreover, such information can dynamically evolve and is frequently obtained from diverse data providers. We claim that an effective management of this diverse information is essential to deliver pertinent recommendations, aligned with user preferences, within the specific context.
However, it is evident from our observation that TRSs are commonly constructed from scratch to suit the specific context in which they operate.
On the contrary, supporting developers with configuration approaches enabling them to easily configure and customize TRSs to their operational environments while preserving commonalities would promote reuse, minimize errors, and accelerate time-to-market. 
In the context of the \textit{RASTA}
project\footnote{\url{http://rasta.unicampania.it/}}, the realization of (hybrid) TRSs offering configuration features plays an important role.
%
In this respect, several approaches have been proposed to facilitate the design, configuration, and deployment of recommender systems in different domains, \eg~e-commerce~\cite{Mettouris2018}, multimedia recommendations~\cite{anelli_elliot_2021}, or complex software systems~\cite{DISIPIO2024101256}. However, offering dedicated utilities to specifically support the development of recommender systems in the travel domain is still challenging.



With the intention of investigating the configuration features offered by the existing systems, and ensuring we do not overlook any pertinent solution, we conduct a systematic mapping study (SMS) by adhering to the well-established guidelines proposed by Petersen~\etal \cite{petersen_guidelines_2015}. Such a process aims to characterize the current landscape of TRSs by analyzing critical aspects for providers \ie the employed algorithm, the data type, the provided travel recommendations, and the configuration support.
Our findings demonstrate that developing TRSs requires the usage of advanced techniques, \ie~standard filtering algorithms are not enough to handle real-time data. More importantly, the limited scenarios covered by the current literature pose critical challenges to the generalizability to the travel domain, \ie most of the approaches are ad-hoc solutions. In this respect, we present possible future directions in supporting TRSs development, intending to address the current challenges derived from our study. 




%% file: src/Motivation.tex

\noindent

In this section, 
we illustrate the significance of configuring TRSs through two exemplary scenarios outlined below.

\noindent
\ding{228} \textbf{Smart City scenario}: In an urban context, several factors must be taken into consideration while planning travel activities. For instance, traffic conditions may cause unwanted delays while moving from a certain POI to another. 
As of today, smart cities are in continuous evolution to also promote their sustainable development, improve human lives, and preserve the environment. They include, among other things, diverse data sources, \eg~mobile devices, sensor networks, open services, and even citizens, potentially providing a plethora of data that might be exploited to better tie travel recommendations. Furthermore, users can modify their preferences at run-time, thus involving additional constraints related to the specific scenario, such as avoiding crowded places or highly polluted areas. 
In this respect, TRSs providers have to carefully take into account those factors while specifying the desired TRS. 
Even though advanced algorithms and hybrid techniques can manage this, providers should analyze the {\em dynamicity} of the context from the design time such that to implement and further deploy configurable TRS.


 
\noindent
\ding{228} \textbf{Natural Park scenario}: Compared to the first scenario, natural parks, \eg~national parks, 
involve a different set of user preferences and contextual information. 
Context-aware factors that are specifically related to the urban scenario are not considered, \eg~traffic levels cannot be measured since vehicles cannot access the park. On the contrary, the presence and conditions of trails, possibly connected with weather conditions, play a crucial role 
since recommendations may involve mostly outdoor activities. Similarly, user profiling should focus on other personal characteristics, \eg~what kind of outdoor activities the user prefers, and their level of expertise in hiking. Furthermore, because of connectivity limitations, data sources like mobile devices or sensor networks may be utilized to a reduced extent, compared to an urban context.
Therefore, a TRS specifically conceived for an urban scenario may not be suitable for recommending travel activities in natural parks. In the worst case, providers are forced to develop the entire system from scratch or re-engineer a consistent part of it. 


Even though context, users, and users' preferences are considered both in the smart city and natural park, each of them poses specific challenges that providers have to handle properly. In this respect, configuring the TRS \textit{a priori} may ease the burden of developing a tailored solution that cannot be adapted to completely different scenarios. In particular, a generic TRS must handle such heterogeneity by devising specific components in its internal design.
Furthermore, we acknowledge the existence of various context-aware TRS. However, we argue that context-awareness does not inherently entail configurability, namely, the possibility to configure specific aspects of TRSs based on the context they are applied. On the contrary, these systems often tend to be rigid and tightly bound to their execution context.

%% file: src/Methodology.tex

In this section, we describe the conducted SMS, 
and 
we discuss the papers' mapping 
to the identified research questions. 

\subsection{Research Methodology} \label{sec:process}
    
To conduct the SMS, we followed guidelines described by Petersen~\etal~\cite{petersen_guidelines_2015}. 
%
Planning and Conducting phases are described in the subsequent paragraphs. The Reporting phase, instead, is discussed in Section~\ref{sec:mapping}. All the data, such as papers dataset, performed classifications, and results, can be found in the online
replication package \cite{replicationpackage}.


\paragraph{\textbf{Planning}}
During this phase, our primary objective is to outline the protocol that will guide our work throughout the conducting phase. 
The starting point consists of the definition of the 
Research Questions~(RQs) we aim to answer. 
%
Specifically, we are interested in addressing the following:
        
        \noindent \ding{228} \rqfirst  
        \noindent
        
        While selecting the recommendation algorithm, providers often reuse off-the-shelf tools and techniques as the most common solution. In certain cases, combining different algorithms may be needed to handle peculiar situations. Thus, we investigate this dimension to understand what are the main adopted algorithms to develop TRSs.
        
        \noindent  \ding{228} \rqsecond 
        \noindent
        
        Once the algorithm has been selected, providers should assess the pertinence of the recommendations according to the context where the system is deployed. 
        Handling heterogeneous input data is a crucial phase that affects the overall process. Thus, we carefully investigate the existing literature to elicit common solutions to cope with this issue. 
        
        \noindent  \ding{228} \rqthird
        \noindent
        
        As said, especially
        providers who are not domain experts need a deep knowledge of the context in which the TRS is supposed to be deployed. Moreover, changing the application context may require reworking the conceived solution from scratch. This research question aims to give an overview of the configuration features currently offered by existing works. 

\smallskip
Once defined, RQs are used to identify keywords to be subsequently employed in search queries within digital libraries.
%
Specifically, we scoured for terms related to this work in the domain of travel recommender systems, narrowing it down to \emph{configurable} (and similar terms) \emph{recommender} (and similar terms) systems for \emph{tourism} (and similar terms). The goal of this query is to demonstrate that while TRSs may include personalization or customization features for the end-user, they do not adequately address the configuration aspect that would assist the providers of these systems. Moreover, when looking for configuration features from the providers' perspective, we cannot completely neglect personalization features that mainly refer to the end-user, since they might be tied. The search includes the titles of works, their abstracts, and keywords. 
Moreover, we limit our query to works published within the past 10 years considering journal and conference papers. Listing \ref{lst:searchString} reports the generic query that has been used to collect existing approaches. To minimize any bias in the process, we ran our final query on three digital libraries: Scopus~\footnote{\url{https://www.scopus.com/}}, IEEE~\footnote{\url{https://ieeexplore.ieee.org/}}, DBLP~\footnote{\url{https://dblp.org/}}.
Here, we would like to clarify that we also utilized the ACM digital library in our search. However, during the query process, we encountered technical problems that were confirmed by the library's technical support team and required time to resolve. 
Thus, we proceeded without incorporating it into our analysis. Notably, we utilized Scopus, which often contains many duplicate results from the ACM digital library.

\begin{lstlisting}  [label=lst:searchString,caption=Search string run on the selected digital libraries.,style=searchstringstyle]
("Configurable" OR "Personali*" OR "Customi*") AND ("Recommender" OR "Advis*" OR "Assist*" OR "Suggest*") AND (("Tourism" OR "eTourism") AND "Travel") AND (PUBYEAR > 2012 AND PUBYEAR < 2024) 
\end{lstlisting}

%

In addition, during the planning phase, we also defined the inclusion and exclusion criteria as seen in Table~\ref{table:criteria}.
        
        \begin{table}[htb!]
        \centering
        \caption{Inclusion and Exclusion Criteria.}
        \begin{tabular}{p{8cm}}
        \hline
        \multicolumn{1}{c}{\textbf{Inclusion criteria}} \\ \hline \hline
        1. Papers that propose TRSs with a certain degree of configuration.     
        \\ \hline
        2. Peer-reviewed papers published in conferences and journals.     
        \\ \hline
        3. Works published within the last 10 years.
        \\ \hline
        \multicolumn{1}{c}{\textbf{Exclusion criteria}} \\ \hline \hline
        1. Papers that propose RS for other application domains.                        \\ \hline
        2. Papers not written in English.                        \\ \hline
        3. Short papers, posters, and tutorials (4-6 pages).                        \\ \hline
        4. Out-of-scope papers, \eg surveys or empirical studies on TRS                    \\ \hline
        \end{tabular}
        \label{table:criteria}
        \end{table}

\paragraph{\textbf{Conducting}}
The retrieving papers process. started with the search query that returned \textbf{203} papers.
The process of removing duplicates excluded \textbf{33} works, bringing the total number 
to \textbf{170}. Subsequently, we selected papers based on their titles, abstracts, and keywords, thus further excluding \textbf{17} works. The final step implies the application of the inclusion and exclusion criteria over \textbf{153} works. 
%
%
We ended up with \textbf{40} primary studies that are reported in Table~\ref{tab:comparison}. For each approach, we elicited four different aspects, aiming to characterize the current trends and identify possible challenges in configuring TRSs, 
described as follows: 

\smallskip

\noindent \ding{228} \textbf{Algorithm:} This feature represents the selected algorithm to perform the recommendations. Depending on the type of the algorithm, we create a categorization based on their high-level functionality. 
We ended up with the following six categories. 
\textit{Collaborative Filtering~(CF)} technique is the most adopted in the RS domain~\cite{Schafer:2007:CFR:1768197.1768208} where input data are typically encoded in user-item matrices that are used to retrieve the outcomes according to different similarity functions. However, this type of algorithm suffers from the so-called cold-start problem, \ie when the system doesn't have an initial set of ratings or items to perform the training phase. In this respect, \textit{content-based~(CB)} \cite{Salter2006CinemaScreenRA} technique provides recommendations to the user based on their previous choices, suggesting items that are similar to their past preferences. \textit{Social-based~(SB)} technique \cite{carrer-neto_social_2012} exploits knowledge graphs, social network data to overcome this limitation. All the abovementioned filtering algorithms can be combined to devise \textit{hybrid-based systems~(HB)}~\cite{burke_hybrid_2002} to handle peculiar situations. With the advent of machine learning, the TRS community has started to consider such models. In our work, we focus on \textit{neural networks~(NN)} and different types of \textit{optimization models~(OP)}. For those algorithms that do not fall under these categories, we labeled them as \textit{OTHER}.


\noindent \ding{228} \textbf{Data Type:} Most of the approaches typically make use of two different types of data \ie historical~(H) and real-time~(RT). The former refers to data that have been collected before the execution of the system. The latter depends on the context and the system may be executed at run-time, \ie on a new set of data. Furthermore, the chosen strategy must be capable of handling heterogeneous data format.

\noindent \ding{228} \textbf{Outcomes:} This feature refers to the final output the systems deliver to users. In particular, a TRS can recommend a ranked list of POIs by computing the similarity considering user preferences or similar POIs. Alternatively, the system can capture the location in chronological order \cite{Noorian2022}, thus providing the so-called itinerary that considers the user's behavior patterns.
Thus, we classify the outcomes of the existing approaches as ranked lists of POIs~(PR) or itineraries~(IT).   

\noindent \ding{228} \textbf{Configuration support:} The main goal of this paper is to investigate to which extent current TRSs can be configured under several aspects. 
Thus, we inspect the collected TRSs from the providers' perspective, 
\eg if some components can be easily deployed without re-implementing the system from scratch or by configuring it.


\begin{table*}[t]
\caption{Analysis of existing approaches}  

\label{tab:comparison}
\centering
\scriptsize

\begin{threeparttable}

\resizebox{\textwidth}{!}{
\begin{tabular}{|l|c|c|c|c|}

\hline
    \textbf{Approach} & 
    \textbf{Algorithm} &    
    \textbf{Data Type} &
    \textbf{Outcomes} &
    \textbf{Configuration support} 
   
\\
\hline \hline

Y. Zhang~\etal~\cite{Y_Zhang2022} & HB & H & PR & User profile and context-aware data  \\ 
\hline 



M. Ummesalma; C. Yashiga~\cite{Ummesalma2021} & HB & H & PR & User preferences and POIs \\
\hline  


J. Neidhardt~\etal~\cite{Neidhardt2014_2} & HB & H & PR  &  User profile \\
\hline

M. E. B. H. Kbaier \etal~\cite{Kbaier2017} & HB & H & PR   & User profile \\
\hline


A. Moreno \etal~\cite{Moreno2013} & HB & H & PR   &  User preferences, demographic info, and context-aware data  \\
\hline

M. Kolahkaj \etal~\cite{Kolahkaj2020} & HB  & RT & PR  & User preferences and POIs  \\
\hline

R. Logesh \etal~\cite{Logesh2018_3}& HB  & H & PR   &  User preferences  \\
\hline

A. Noorian \etal~\cite{Noorian2022} & HB & H & IT  & User preferences \\
\hline


L. Esmaeili~\cite{Esmaeili2020} & HB & H & IT & User preferences  \\
\hline  

T. D. Pessemier \etal~\cite{Pessemier2016} & HB & H & PR  &  User preferences \\
\hline

R. Logesh \etal~\cite{Logesh2018_2}& HB & H & PR  & User preferences \\
\hline



I. Y. Choi~\etal~\cite{Choi2015} & HB & H & IT & User profile and preferences \\
\hline





K. I. Kotsopoulos~\etal~\cite{Kotsopoulos2022} & HB & RT & IT & User preferences, itineraries, and POIs \\
\hline




Q. Gu~\etal~\cite{Gu2019} & HB & H & PR & User preferences and flight features \\ 
\hline







J. Xu \etal~\cite{Xu2021} &  NN   & RT & IT  & Users (profile, itinerary, and behavior sequence) and POIs    \\ 
\hline


J. Ye \etal~\cite{Ye2019} & NN  & H & PR  & User preferences and POIs \\ 
\hline

L. Chen \etal~\cite{LChen2022}& NN & RT & PR   &  User preferences \\ 
\hline 

L. Chen \etal~\cite{LChen2023} & NN & H & PR   &  User preferences  \\ 

\hline

G. Zhu \etal~\cite{Zhu2021} & NN & H & PR &  User preferences  \\  
\hline

L. Chen \etal~\cite{LChen2020} & OP  & RT & PR  &  User preferences \\ 
\hline


C. Tan \etal~\cite{Tan2014}& OP & RT & IT & User features \\
\hline 

J. L. Sarkar~\etal~\cite{Sarkar2021} & OP & H & IT & User preferences and POIs \\ 
\hline

E. Streviniotis; G. Chalkiadakis~\cite{Streviniotis2022}& OP & H & PR  & User and POIs\\
\hline  


F. Teklemicael \etal~\cite{Teklemicael2016} & CF  & H & PR  & User preferences  \\
\hline


Q. Miao \etal~\cite{Miao2020} & CF & H & PR   &  Users and POIs\\
\hline

F. Leal \etal~\cite{Leal2018} & CF  & H & PR  & User profile \\
\hline

Q. Qi~\etal~\cite{Qi2018} & CF & H & PR  & User preferences, social attributes, and consumption behavior \\
\hline








S. Missaoui~\etal~\cite{Missaoui2019} & CB & H & PR  &  User preferences \\
\hline

Z. Bahramian; R. Abbaspoura~\cite{Bahramian2015} & CB & H & PR &  User profile  \\
\hline  



M. Figueredo  \etal~\cite{Figueiredo2018} & SB & H & PR  & User preferences \\
\hline



L. Ravi \etal~\cite{Ravi2019} & SB & H & PR  &  User profile and POIs\\ 
\hline


J. Shen \etal~\cite{Shen2016}  & OTHER & H & PR  &  User preferences  \\ 
\hline

X. Shao \etal~\cite{Shao2019} & OTHER & H & PR  & User preferences  \\ 
\hline

C. Bin \etal~\cite{Bin2019} & OTHER & H & PR  & User preferences and POIs  \\ 
\hline

S. Nikookar~\etal~\cite{Nikookar2022} & OTHER & H  & IT & User preferences and POIs \\ 
\hline




T. Alenezi; L. Hirtle~\cite{Alenezi2022} & OTHER & RT & PR & User preferences, POIs  \\
\hline 


H. Chiang; T. Huang~\cite{Chiang2015} & OTHER & RT & PR  &  User preferences and POIs\\
\hline 

I. Benouaret; D. Lenne~\cite{Benouaret2017} & OTHER & H & PR  &  User preferences and POIs \\ \hline 


A. Gehlot; R. Singh~\cite{Gehlot2022}& OTHER & H & IT  & User preferences  \\ 
\hline  

T. Huang~\etal~\cite{Huang2020} & OTHER & H & IT & User preferences and itineraries \\ 
\hline



\end{tabular}}

\end{threeparttable}
\end{table*}


\noindent

\subsection{Mapping of papers} \label{sec:mapping}

In this section, we discuss the results obtained by the SMS by mapping the analyzed papers to the three stated RQs. 


\subsubsection{\rqfirst} \label{sec:rq1}

To answer this question, we carefully analyzed the underpinning algorithm used in the approaches presented in Table \ref{tab:comparison}. Discussing in depth each algorithm presented in the selected approach is out of the scope of this paper. Instead, we focus on identifying \textit{i)} the main techniques used to implement the TRS and \textit{ii)} the type of travel recommendations that they support. As shown in Table \ref{tab:comparison}, we categorize them into seven different high-level categories. 

It is worth noting that the most used in the TRS domain is the hybrid-based (HB) approach \cite{burke_hybrid_2002}, \ie~\textbf{14} of the approaches employ this technique to perform the recommendation. Such a result can be explained by the fact that the HB algorithm involves a combination of traditional filtering algorithms. For instance, Noorian~\etal \cite{Noorian2022} combine social information and knowledge-based systems \cite{10.1145/3269206.3271739} with collaborative filtering \cite{Schafer:2007:CFR:1768197.1768208} and context-aware information to provide optimized tourist itineraries. Kotsopoulos \etal~\cite{Kotsopoulos2022} confirm that HB outperforms CF and CB algorithms in terms of accuracy. \textbf{5} of the elicited approaches employ different neural network (NN) models, \ie long-short term memory (LSTM), graph neural network (GNN), and deep neural network (DNN). Notably, Zhu \etal \cite{Zhu2021} and Ye \etal \cite{Ye2019} report that LSTMs can effectively capture the behavior of the users over time, thus improving traditional session-based approaches~\cite{jannach2022session} in which the users are forced to provide feedback during different training sessions. Among the approaches that adopt optimization strategies (OP), two out of four papers make use of the well-founded Bayesian network~(BN) model, namely Streviniotis and Chalkiadakis \cite{Streviniotis2022} and Tan \etal~\cite{Tan2014}. 
Concerning the traditional approaches, the most used is collaborative filtering (CF), with \textbf{4} approaches, followed by content-based~(CB) and social-based~(SB), both with \textbf{2} approaches. Such a lower number can be explained by the fact that those algorithms have been combined in the HB systems as discussed previously. Finally, the conducted SMS highlights the presence of \textbf{9} papers that adopt algorithms or heuristic strategies that cannot be categorized in the previous six categories. Among them, Huang~\etal~\cite{Huang2020} employs the niching genetic algorithm (NGA) to solve the multi-itinerary planning problem. 
Chiang and Haung~\cite{Chiang2015} focus on implementing personalized travel planning based on time scheduling. 
From the conducted analysis, we can observe that TRSs require advanced techniques, or a combination of well-founded algorithms, to recommend pertinent items to users. 
This observation is confirmed by analyzing the outcomes provided by the selected TRSs. In particular, most of the systems suggest ranked POIs~(PR) rather than specific itineraries~(IT), \ie 87\% and 13\% of the approaches, respectively. Furthermore, planning a complete itinerary that embodies user preferences and context-aware information requires dedicated solutions, \ie~the class of algorithms employed are the HB, NN, and OP. Consequently, traditional filtering techniques are limited to providing only similar POIs to the final user.

\begin{shadedbox}
\textbf{Answer to RQ$_1$:} Altogether, the most spread algorithm is the hybrid-based one. Providers tend to combine existing techniques rather than developing new models from scratch, especially when recommending itineraries.
\end{shadedbox}


\subsubsection{\rqsecond}

To answer this question, we discuss the kind of data mostly used in the TRSs approaches identified in Section \ref{sec:process}, \ie~historical~(H) and real-time~(RT) data. While we do not focus on data structure, we focus on better characterizing the collected papers in terms of \textit{i)} what are the most used information considering the application context and \textit{ii)} how the identified approaches manage the dynamic context information to update the recommendation items accordingly. By looking at Table \ref{tab:comparison}, it is evident that the majority of the analyzed TRSs do not manage RT data, \ie only nine works out of forty exploit them to perform the travel recommendations. By focusing on those approaches, we report that the majority of them consider real-time data related to users rather than the context itself. For instance, in this respect, Kotsopoulos \etal~\cite{Kotsopoulos2022} dynamically update the user's location during cruise travel to adjust the recommendation at run-time. Similarly, Alenezi and Hirtle \cite{Alenezi2022} collect the attractions visited by the users while the system is running to adjust the recommendation compared to similar travelers. On the contrary, only a few approaches move a step forward by analyzing also further data than users' behavior. In this respect, Xu \etal \cite{Xu2021} combine the duration time of the travel with the user behavior sequence while Chiang and Haung \cite{Chiang2015} consider travel delays in their custom scheduling optimization model. Concerning the encoding technique, our investigation reveals that the topics modeling approach and attention layers are the most adopted techniques by the providers.  

\begin{figure}[htb!]
    \centering
    \includegraphics[width=0.9\linewidth]{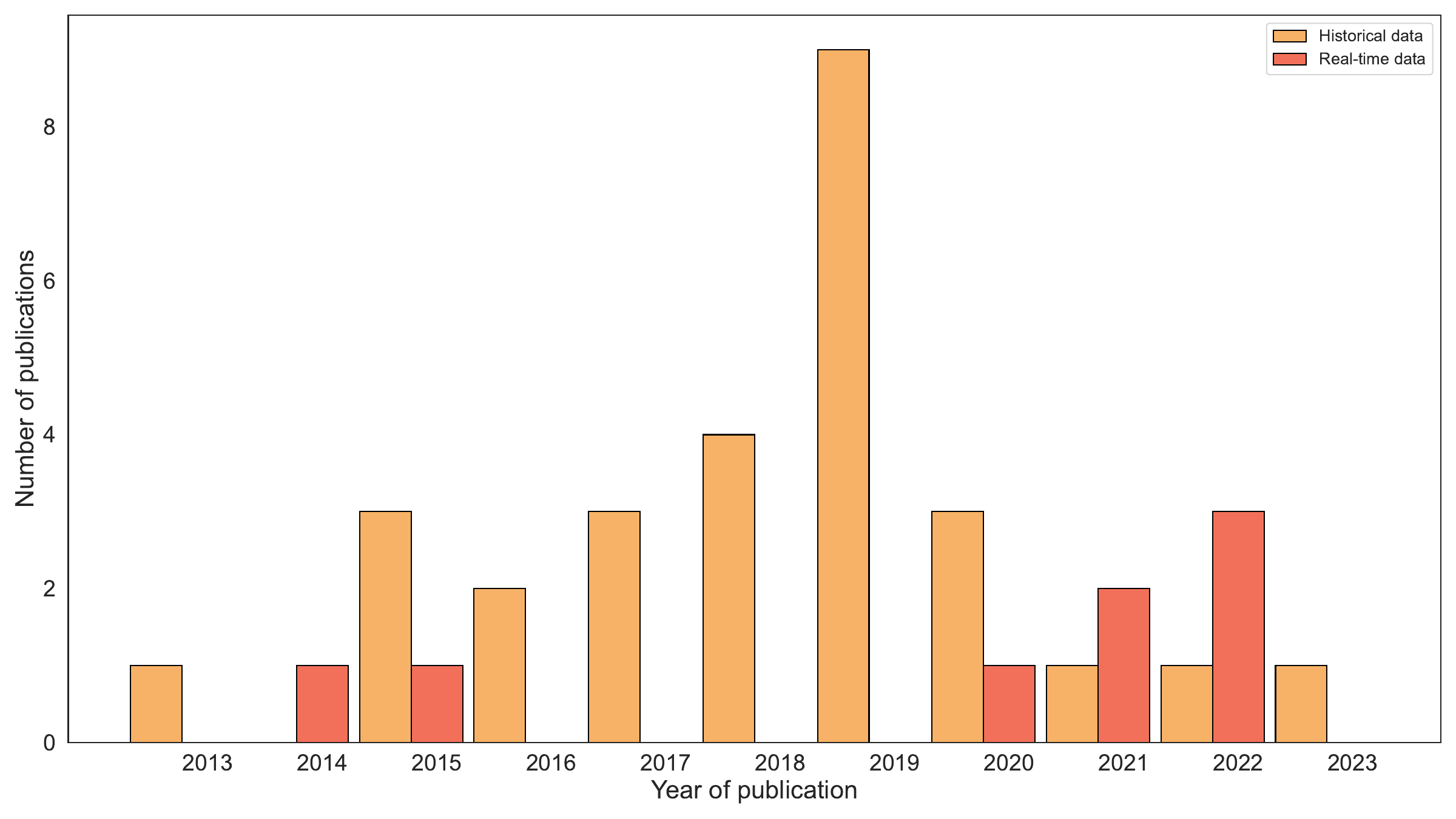}
    \caption{Evolution of the data types used in the considered papers.} 
    \label{fig:rq2_plot}
\end{figure}

To further investigate the usage of RT data, we measured the distribution of approaches that consider real-time data through the selected time window, \ie 2013-2023. As shown in Figure~\ref{fig:rq2_plot}, real-time data usage has increased throughout the years. In particular, approaches proposed by Tan~\etal~\cite{Tan2014} and Chiang and Haung \cite{Chiang2015} are the prior works that employ real-time data in their recommendations in 2014 and 2015, respectively. However, we witnessed that the usage of real-time data has grown in recent years, \ie spanning from 2020 to  2022 the number of TRSs trained with RT data is increasing. 
No papers considered RT data in 2023; however this might be because the query was conducted in January 2024.
%
However, 
most RT data refer to the users, \eg location or behavior patterns. 
In the analyzed approaches, time duration is the most considered context-related RT data; however, none of them consider the unique characteristics of the application scenario.

\begin{shadedbox}
\textbf{Answer to RQ$_2$:} 
Existing TRSs tend to make use of historical data encoded in a static knowledge base. Furthermore, it seems that traditional filtering techniques are not enough to handle dynamic data, \ie the examined TRS may not support the dynamicity of the context even though this aspect is becoming relevant.
 \label{sec:rq2}
\end{shadedbox}





\subsubsection{\rqthird} \label{sec:rq3}

This question discusses the level of configuration support offered by the identified approaches. To this end, we look at the whole system by analyzing which features can be configured by the user or the provider. 
In particular, we identified three main types of configuration dimensions, \ie user preferences, POIs, and user profiles. In this study, we consider user preferences as the elements that the user can express or configure directly in the system, even at run-time. 
The configurable elements that are bounded by a specific domain or peculiar have been categorized as \emph{Other}. 

As expected, user preferences appear in most of the works as a configurable component of the system, \ie~the underpinning algorithm needs to adapt its outcomes according to different users' constraints. However, the presence of this configuration can be expressed in different ways according to the TRS application scenario or the chosen algorithm. For instance, Kbaier \etal \cite{Kbaier2017} exploit the activity category to recommend similar locations. Zhang \etal \cite{Y_Zhang2022} collect the user preferences to find similar passengers to support air traveling. Moreover, user preference can be induced using user profiling or by capturing implicit feedback, thus not requiring direct action. Logesh \etal \cite{Logesh2018_3} collect activities and preferences of single users to aggregate them and recommend POIs for groups. Leal \etal \cite{Leal2018} exploit users' reputations to build a collaborative framework based on trustworthiness.
Other approaches, such as Ravi \etal \cite{Ravi2019} and Figueredo \etal \cite{Figueiredo2018}, conduct the profiling activities using interaction with social media. 
Finally, user preferences can be used to modify the list of POIs according to time scheduling preferences as shown in Xu \cite{Xu2021} or consider the preferred season as done by Bin \etal \cite{Bin2019}.

\begin{shadedbox}
\textbf{Answer to RQ$_3$:}
Altogether, we see a lack of configuration support for providers in the whole process, \ie from the design to the actual deployment. Existing TRSs are focused on customizing the travel experience on top of users' preferences. 
None of the approaches provide configuration features given the application scenario.


\end{shadedbox}


%% file: src/Discussion.tex

This section aims to discuss the findings of our study by identifying existing challenges and takeaways. 
Concerning the algorithm used by TRSs, we observed that tailored solutions have to be developed to handle peculiar situations, \ie multi-itinerary planning or considering user behavior at runtime. However, none of the collected approaches facilitate providers to easily configure these advanced algorithms for various contexts where they may be applied. 
For instance, deploying a collaborative filtering algorithm in the natural park scenario (see Section \ref{sec:Motivation}) can be challenging, as this kind of scenario offers real-time data that cannot be easily encoded as user-item matrices, \eg the actual condition of the trails cannot be predicted using solely historical data.

One solution is combining standard-filtering techniques to build a hybrid-based approach, even though the performance of the system may not be the same by changing the application context. In this respect, most of the analyzed approaches support travel recommendations in an urban context. 
Interestingly, none of the examined papers provide a TRS in a context 
similar to the natural park scenario. 
Our intuition is that this type of context presents a set of challenges that are not manageable using traditional filtering approaches since it involves dynamic conditions alongside historical data.
For instance, sudden changes in weather conditions can impact the recommendation in the natural park scenario. Additionally, users may not access weather forecasting due to the limited network connection. 
Hence, supporting configuration in selecting the proper algorithm given the application scenario can facilitate the providers in the design of their system.


%


\textbf{Takeaway \#1:} 
\textit{Our study demonstrates that mapping the recommendation algorithm to the application scenario is a daunting task, especially considering the dynamicity of certain scenarios, \eg natural parks. Thus, providers need support in the configuration and deployment of the 
algorithm to \textit{i)} handle context-aware dynamicity and \textit{ii)} cover novel 
scenarios.}


A further critical aspect is the handling of real-time data, as they are becoming predominant. 
While traditional algorithms perform well with historical data, the dynamicity of the context requires techniques that often depend on the chosen model deployed. 
Leisure activities in urban areas may suffer from transportation disruptions, such as strikes. Similarly, trails in natural parks can become inaccessible due to adverse weather or natural events like avalanches.

Strategies like topic modeling or attention layers in neural networks seem to be effective, even though they require deep knowledge of the underpinning algorithm to be effectively employed. Moreover, none of the examined papers discloses the necessary data to replicate the obtained results and input data may vary over time. 
%
In this respect, eliciting the proper building blocks to run and execute the 
system for non-expert providers is challenging.


\textbf{Takeaway \#2:}
\textit{Our study underlines the importance of the real-time nature of input data during the development of TRSs even though few approaches is capable of handling them properly. Moreover, data related to the application context is sometimes neglected by some approaches 
focused on the end-user preferences. Therefore, providers could benefit from a dedicated framework that facilitates the configuration of data.}



A possible solution may rely on the idea of engineering software product lines (SPL) \cite{10.5555/501065} and family of products~\cite{van2010software}, \eg~TRSs that share similarities and, at the same time, need customization to be tied to the execution contexts. Ideally, providers may engineer TRSs by sharing a common production methodology while configuring data structures, algorithm hyper-parameters, and peculiar elements of the application context.
Specifically, providers can make use of SPLs to easily configure the system by adhering to the Feature-Oriented Software Development (FOSD) paradigm \cite{thum2014featureide}. 
Such a methodology exploits \textit{feature models} where the developer can specify all the desired product characteristics before deploying the actual system. 
Recent studies demonstrate that feature models succeed in facilitating software developers in specifying complex software systems, \ie recommender system for developers \cite{DISIPIO2024101256} and machine learning pipeline \cite{d2023democratizing}. Besides the feature models, low-code development platforms (LCDPs) may help in configuring TRSs, as they provide complete support for non-expert users to develop software applications \cite{forsyth_top_2021}. In particular, providers can make use of general-purpose LCPD to train and test the algorithm using historical data, \eg POIs location in smart cities or trails difficulty in natural parks. Even though some LCPDs, \eg NodeRed \footnote{\url{https://nodered.org/}} are in place to design and test event-driven applications, there is no dedicated solution for developing TRSs using this paradigm.
Furthermore, recent studies demonstrate the usage of Large Language Models (LLMs) \cite{WONG2023253} or dedicated chatbots \cite{BENADDI2024275} that can be used to customize several aspects of travel experiences for the users. Our intuition is that providers can employ these cutting-edge technologies to develop custom TRSs even though an in-depth analysis is required. 


\textbf{Takeaway \#3:}
\textit{Providers might make use of software product line engineering, low-code platforms, or specialized AI-based tools that guide the design-to-deployment process. However, using those technologies requires a deep knowledge to handle their inner limitations, \eg issues of scalability in software product line engineering, limited customization in low-code platforms, or potential biases in AI-based tools. 
}




In summary, our study reveals a lack of a common methodology for configuring critical TRS components easily \eg~eliciting suitable algorithms or neglecting context-aware information. Recent efforts tackle this issue, but implementing tailored solutions still requires significant effort to provide timely travel recommendations.

%% file: src/Threats.tex
This section discusses threats to validity that may hamper the outcomes of the study \cite{wohlin_experimentation_2012}. 
%
\textit{Construct validity} is related to the process followed to collect and analyze the current literature. Our study may miss relevant approaches in the TRS domain, also due to the problem faced with ACM, and the selection phase might be biased. 
To mitigate the former, we followed well-established guidelines to perform the SMS, by considering three different digital libraries. Moreover, as regards the ACM digital library, we utilized Scopus, which typically contains many duplicate results from ACM. Concerning the latter, we created two independent groups of authors, \ie the first two authors collected and analyzed the papers to extract the features discussed in Table \ref{tab:comparison} and the last two revised the table, through multiple iterations, searching and solving possible misalignment. 
Concerning \textit{internal validity}, \ie possible threats to the employed methodology, the selection of the keywords for querying the digital libraries may lead to an incomplete landscape of the problem, \ie missing relevant papers. 
To cope with this, we carefully design the used search string by considering synonyms, 
and we exploit the advanced search feature of the three search engines. 
%
\textit{External validity} concerns the generalizability of the study \ie to what extent the findings can be generalized inside or outside of the considered population. To address this, we rigorously adhere to the SMS guidelines \cite{petersen_guidelines_2015}. Furthermore, we manually extract and compare the selected approaches in terms of relevant features for the providers. While we narrowed down the scope to assess the configuration features offered by existing TRSs, the identified features are generic enough to be applied to additional classes of recommender systems. 
%
\textit{Conclusion validity} represents threats that could affect the relation between the study and the observed findings. 
We carefully read all the selected approaches, 
and we report relevant descriptive statistics to better represent the population.

%% file: src/Conclusion.tex
TRSs bring additional challenges for providers compared to traditional recommender systems, \ie the managing of real-time data and the heterogeneity of application scenarios. To investigate this, we conduct a systematic mapping study by analyzing the existing landscape in the domain.
By analyzing the identified primary studies, we discovered that none of the resulting approaches offer capabilities to configure a TRS since current approaches resemble off-the-shelf solutions. To handle this, we envision possible future directions 
to address the identified challenges. 
%
As future work, we plan to extend the conducted study by considering further aspects of TRSs, \eg the evaluation of the provided travel recommendations. In addition, we will investigate the techniques mentioned in the discussion section to devise a generic framework capable of facilitating the providers' development activities. 

%% file: main.bbl
\begin{thebibliography}{10}
\providecommand{\url}[1]{#1}
\csname url@samestyle\endcsname
\providecommand{\newblock}{\relax}
\providecommand{\bibinfo}[2]{#2}
\providecommand{\BIBentrySTDinterwordspacing}{\spaceskip=0pt\relax}
\providecommand{\BIBentryALTinterwordstretchfactor}{4}
\providecommand{\BIBentryALTinterwordspacing}{\spaceskip=\fontdimen2\font plus
\BIBentryALTinterwordstretchfactor\fontdimen3\font minus \fontdimen4\font\relax}
\providecommand{\BIBforeignlanguage}[2]{{%
\expandafter\ifx\csname l@#1\endcsname\relax
\typeout{** WARNING: IEEEtran.bst: No hyphenation pattern has been}%
\typeout{** loaded for the language `#1'. Using the pattern for}%
\typeout{** the default language instead.}%
\else
\language=\csname l@#1\endcsname
\fi
#2}}
\providecommand{\BIBdecl}{\relax}
\BIBdecl

\bibitem{hamid_how_2021}
R.~A. Hamid, A.~S. Albahri, J.~K. Alwan, Z.~T. Al-qaysi, O.~S. Albahri, A.~A. Zaidan, A.~Alnoor, A.~H. Alamoodi, and B.~B. Zaidan, ``How smart is e-tourism? {A} systematic review of smart tourism recommendation system applying data management,'' \emph{Computer Science Review}, 2021.

\bibitem{delic_research_nodate}
A.~Delic, J.~Neidhardt, N.~Nguyen, and F.~Ricci, ``Research methods for group recommender system,'' in \emph{RecTour 2016: Workshop on Recommenders in Tourism, RecSys 2016}, vol. 1685.\hskip 1em plus 0.5em minus 0.4em\relax CEUR, pp. 30--37.

\bibitem{a9d77a28-8be9-31fc-842d-09259c364c5f}
E.~COHEN, ``Toward a sociology of international tourism,'' \emph{Social Research}, vol.~39, no.~1, pp. 164--182, 1972.

\bibitem{ricci_travel_2002}
\BIBentryALTinterwordspacing
F.~Ricci, ``Travel recommender systems,'' 2002. [Online]. Available: \url{https://api.semanticscholar.org/CorpusID:15595894}
\BIBentrySTDinterwordspacing

\bibitem{Ricci2011}
F.~Ricci, L.~Rokach, and B.~Shapira, \emph{Introduction to Recommender Systems Handbook}.\hskip 1em plus 0.5em minus 0.4em\relax Springer US, 2011, pp. 1--35.

\bibitem{Tkalcic2015}
M.~Tkalcic and L.~Chen, \emph{Personality and Recommender Systems}.\hskip 1em plus 0.5em minus 0.4em\relax Springer US, 2015, pp. 715--739.

\bibitem{Mettouris2018}
C.~Mettouris, A.~Achilleos, G.~Kapitsaki, and G.~A. Papadopoulos, \emph{The {UbiCARS} {Model}-{Driven} {Framework}: {Automating} {Development} of {Recommender} {Systems} for {Commerce}}, ser. Lecture {Notes} in {Computer} {Science}.\hskip 1em plus 0.5em minus 0.4em\relax Springer International Publishing, 2018, pp. 37--53.

\bibitem{anelli_elliot_2021}
V.~W. Anelli, A.~Bellogin, A.~Ferrara, D.~Malitesta, F.~A. Merra, C.~Pomo, F.~M. Donini, and T.~Di~Noia, ``Elliot: A comprehensive and rigorous framework for reproducible recommender systems evaluation,'' New York, NY, USA, p. 2405–2414, 2021.

\bibitem{DISIPIO2024101256}
C.~{Di Sipio}, J.~{Di Rocco}, D.~{Di Ruscio}, and P.~T. Nguyen, ``Lev4rec: A feature-based approach to engineering rsses,'' \emph{Journal of Computer Languages}, vol.~78, p. 101256, 2024.

\bibitem{petersen_guidelines_2015}
K.~Petersen, S.~Vakkalanka, and L.~Kuzniarz, ``\BIBforeignlanguage{en}{Guidelines for conducting systematic mapping studies in software engineering: {An} update},'' \emph{\BIBforeignlanguage{en}{Information and Software Technology}}, vol.~64, pp. 1--18, Aug. 2015.

\bibitem{replicationpackage}
\BIBentryALTinterwordspacing
R.~{Simioni Pereira}, C.~Di~Sipio, M.~De~Sanctis, and L.~Iovino, ``Replication package for this study,'' May 2024. [Online]. Available: \url{https://github.com/ricksonsimioni/On-the-need-for-configurable-travel-recommender-systems-A-systematic-mapping-study}
\BIBentrySTDinterwordspacing

\bibitem{Schafer:2007:CFR:1768197.1768208}
J.~B. Schafer, D.~Frankowski, J.~Herlocker, and S.~Sen, \emph{Collaborative Filtering Recommender Systems}.\hskip 1em plus 0.5em minus 0.4em\relax Springer Berlin Heidelberg, 2007, pp. 291--324.

\bibitem{Salter2006CinemaScreenRA}
J.~Salter and N.~Antonopoulos, ``Cinemascreen recommender agent: combining collaborative and content-based filtering,'' \emph{IEEE Intelligent Systems}, vol.~21, pp. 35--41, 2006.

\bibitem{carrer-neto_social_2012}
W.~Carrer-Neto, M.~L. Hern\'{a}ndez-Alcaraz, R.~Valencia-Garc\'{\i}a, and F.~Garc\'{\i}a-S\'{a}nchez, ``Social knowledge-based recommender system. {Application} to the movies domain,'' \emph{Expert Systems with Applications}, vol.~39, no.~12, pp. 10\,990--11\,000, 2012.

\bibitem{burke_hybrid_2002}
R.~Burke, ``Hybrid {Recommender} {Systems}: {Survey} and {Experiments},'' \emph{User Modeling and User-Adapted Interaction}, vol.~12, no.~4, pp. 331--370, 2002.

\bibitem{Noorian2022}
A.~Noorian, A.~Harounabadi, and R.~Ravanmehr, ``A novel sequence-aware personalized recommendation system based on multidimensional information,'' \emph{Expert Systems with Applications}, vol. 202, p. 117079, Sep. 2022.

\bibitem{Y_Zhang2022}
Y.~Zhang, W.~Luo, M.~Li, and T.~Chen, ``Contextualized recommendation of aviation ancillary services based on passenger portraits,'' \emph{IEEE Transactions on Aerospace and Electronic Systems}, vol.~58, no.~6, p. 5078–5088, Dec. 2022.

\bibitem{Ummesalma2021}
U.~M and Y.~C, ``{COLPOUSIT}: A hybrid model for tourist place recommendation based on machine learning algorithms,'' in \emph{2021 5th International Conference on Trends in Electronics and Informatics ({ICOEI})}.\hskip 1em plus 0.5em minus 0.4em\relax {IEEE}, Jun. 2021.

\bibitem{Neidhardt2014_2}
J.~Neidhardt, L.~Seyfang, R.~Schuster, and H.~Werthner, ``A picture-based approach to recommender systems,'' \emph{Information Technology \&amp; Tourism}, vol.~15, no.~1, p. 49–69, Sep. 2014.

\bibitem{Kbaier2017}
M.~E. B.~H. Kbaier, H.~Masri, and S.~Krichen, ``A personalized hybrid tourism recommender system,'' in \emph{2017 IEEE/ACS 14th International Conference on Computer Systems and Applications (AICCSA)}, 2017, pp. 244--250.

\bibitem{Moreno2013}
A.~Moreno, A.~Valls, D.~Isern, L.~Marin, and J.~Borr\`{a}s, ``Sigtur/e-destination: Ontology-based personalized recommendation of tourism and leisure activities,'' \emph{Engineering Applications of Artificial Intelligence}, vol.~26, no.~1, p. 633–651, Jan. 2013.

\bibitem{Kolahkaj2020}
M.~Kolahkaj, A.~Harounabadi, A.~Nikravanshalmani, and R.~Chinipardaz, ``A hybrid context-aware approach for e-tourism package recommendation based on asymmetric similarity measurement and sequential pattern mining,'' \emph{Electronic Commerce Research and Applications}, vol.~42, p. 100978, Jul. 2020.

\bibitem{Logesh2018_3}
R.~Logesh, V.~Subramaniyaswamy, V.~Vijayakumar, and X.~Li, ``Efficient user profiling based intelligent travel recommender system for individual and group of users,'' \emph{Mobile Networks and Applications}, vol.~24, no.~3, p. 1018–1033, May 2018.

\bibitem{Esmaeili2020}
L.~Esmaeili, S.~Mardani, S.~A.~H. Golpayegani, and Z.~Z. Madar, ``A novel tourism recommender system in the context of social commerce,'' \emph{Expert Systems with Applications}, vol. 149, p. 113301, Jul. 2020.

\bibitem{Pessemier2016}
T.~D. Pessemier, J.~Dhondt, and L.~Martens, ``Hybrid group recommendations for a travel service,'' \emph{Multimedia Tools and Applications}, vol.~76, no.~2, p. 2787–2811, Jan. 2016.

\bibitem{Logesh2018_2}
R.~Logesh, V.~Subramaniyaswamy, and V.~Vijayakumar, ``A personalised travel recommender system utilising social network profile and accurate gps data,'' \emph{Electronic Government, an International Journal}, vol.~14, no.~1, p.~90, 2018.

\bibitem{Choi2015}
I.~Y. Choi, J.~K. Kim, and Y.~U. Ryu, ``A two-tiered recommender system for tourism product recommendations,'' in \emph{2015 48th Hawaii International Conference on System Sciences}.\hskip 1em plus 0.5em minus 0.4em\relax IEEE, Jan. 2015.

\bibitem{Kotsopoulos2022}
K.~I. Kotsopoulos, G.~Pavlidis, E.~Viennas, E.~Baratis, G.~Giannopoulou, A.~Papadopoulos, M.~Spilios, P.~Alexogianni, and E.~Sakkopoulos, ``Happycruise - integrated information system of personalized information and security in the tourism industry,'' in \emph{2022 13th International Conference on Information, Intelligence, Systems amp; Applications (IISA)}.\hskip 1em plus 0.5em minus 0.4em\relax IEEE, 2022.

\bibitem{Gu2019}
Q.~Gu, J.~Cao, Y.~Zhao, and Y.~Tan, ``Addressing the cold-start problem in personalized flight ticket recommendation,'' \emph{IEEE Access}, vol.~7, p. 67178–67189, 2019.

\bibitem{Xu2021}
J.~Xu, Z.~Wang, Z.~Chen, D.~Lv, Y.~Yu, and C.~Xu, ``Itinerary-aware personalized deep matching at fliggy,'' in \emph{Proceedings of the Web Conference 2021}, ser. WWW '21.\hskip 1em plus 0.5em minus 0.4em\relax Association for Computing Machinery, 2021, p. 3234–3245.

\bibitem{Ye2019}
J.~Ye, Q.~Xiong, Q.~Li, M.~Gao, and R.~Xu, ``Tourism service recommendation based on user influence in social networks and time series,'' in \emph{2019 IEEE 21st International Conference on High Performance Computing and Communications; IEEE 17th International Conference on Smart City; IEEE 5th International Conference on Data Science and Systems (HPCC/SmartCity/DSS)}, 2019, pp. 1445--1451.

\bibitem{LChen2022}
L.~Chen, J.~Cao, Y.~Wang, W.~Liang, and G.~Zhu, ``Multi-view graph attention network for travel recommendation,'' \emph{Expert Systems with Applications}, vol. 191, p. 116234, Apr. 2022.

\bibitem{LChen2023}
L.~Chen, J.~Cao, H.~Tao, and J.~Wu, ``Trip reinforcement recommendation with graph-based representation learning,'' \emph{ACM Transactions on Knowledge Discovery from Data}, vol.~17, no.~4, p. 1–20, Feb. 2023.

\bibitem{Zhu2021}
G.~Zhu, Y.~Wang, J.~Cao, Z.~Bu, S.~Yang, W.~Liang, and J.~Liu, ``Neural attentive travel package recommendation via exploiting long-term and short-term behaviors,'' \emph{Knowledge-Based Systems}, vol. 211, p. 106511, Jan. 2021.

\bibitem{LChen2020}
L.~Chen, Z.~Wu, J.~Cao, G.~Zhu, and Y.~Ge, ``Travel recommendation via fusing multi-auxiliary information into matrix factorization,'' \emph{ACM Transactions on Intelligent Systems and Technology}, vol.~11, no.~2, p. 1–24, Jan. 2020.

\bibitem{Tan2014}
C.~Tan, Q.~Liu, E.~Chen, H.~Xiong, and X.~Wu, ``Object-oriented travel package recommendation,'' \emph{ACM Transactions on Intelligent Systems and Technology}, vol.~5, no.~3, p. 1–26, Sep. 2014.

\bibitem{Sarkar2021}
J.~L. Sarkar, A.~Majumder, C.~R. Panigrahi, V.~Ramasamy, and R.~Mall, ``{TRIPTOUR}:a multi-itinerary tourist recommendation engine based on poi visits interval,'' in \emph{2021 12th International Conference on Computing Communication and Networking Technologies (ICCCNT)}.\hskip 1em plus 0.5em minus 0.4em\relax IEEE, 2021.

\bibitem{Streviniotis2022}
E.~Streviniotis and G.~Chalkiadakis, ``Multiwinner election mechanisms for diverse personalized bayesian recommendations for the tourism domain,'' in \emph{Proceedings of the 2022 Workshop on Recommenders in Tourism, RecTour, Seattle, WA, USA}, vol.~22, 2022.

\bibitem{Teklemicael2016}
F.~Teklemicael, Y.~Zhang, Y.~Wu, Y.~Yin, and C.~Xing, \emph{Toward Gamified Personality Acquisition in Travel Recommender Systems}.\hskip 1em plus 0.5em minus 0.4em\relax Springer International Publishing, 2016, p. 375–385.

\bibitem{Miao2020}
Q.~Miao, L.~Wu, and J.~Yang, ``Classification of tourism english talents based on relevant features mining and information fusion,'' in \emph{Lecture Notes in Computer Science}.\hskip 1em plus 0.5em minus 0.4em\relax Springer International Publishing, 2020, pp. 664--673.

\bibitem{Leal2018}
F.~Leal, B.~Malheiro, and J.~C. Burguillo, ``Trust and reputation modelling for tourism recommendations supported by crowdsourcing,'' in \emph{Advances in Intelligent Systems and Computing}.\hskip 1em plus 0.5em minus 0.4em\relax Springer International Publishing, 2018, pp. 829--838.

\bibitem{Qi2018}
Q.~Qi, J.~Cao, Y.~Tan, and Q.~Xiao, ``Cross-domain recommendation method in tourism,'' in \emph{2018 IEEE International Conference on Progress in Informatics and Computing (PIC)}.\hskip 1em plus 0.5em minus 0.4em\relax IEEE, Dec. 2018.

\bibitem{Missaoui2019}
S.~Missaoui, F.~Kassem, M.~Viviani, A.~Agostini, R.~Faiz, and G.~Pasi, ``Looker: a mobile, personalized recommender system in the tourism domain based on social media user-generated content,'' \emph{Personal and Ubiquitous Computing}, vol.~23, no.~2, p. 181–197, Jan. 2019.

\bibitem{Bahramian2015}
Z.~Bahramian and R.~Ali~Abbaspour, ``An ontology-based tourism recommender system based on spreading activation model,'' \emph{The International Archives of the Photogrammetry, Remote Sensing and Spatial Information Sciences}, vol. XL-1/W5, p. 83–90, Dec. 2015.

\bibitem{Figueiredo2018}
M.~Figueredo, J.~Ribeiro, N.~Cacho, A.~Thome, A.~Cacho, F.~Lopes, and V.~Araujo, ``From photos to travel itinerary: A tourism recommender system for smart tourism destination,'' in \emph{2018 IEEE Fourth International Conference on Big Data Computing Service and Applications (BigDataService)}, 2018, pp. 85--92.

\bibitem{Ravi2019}
L.~Ravi, V.~Subramaniyaswamy, V.~Vijayakumar, S.~Chen, A.~Karmel, and M.~Devarajan, ``Hybrid location-based recommender system for mobility and travel planning,'' \emph{Mobile Networks and Applications}, vol.~24, no.~4, p. 1226–1239, Apr. 2019.

\bibitem{Shen2016}
J.~Shen, C.~Deng, and X.~Gao, ``Attraction recommendation: Towards personalized tourism via collective intelligence,'' \emph{Neurocomputing}, vol. 173, p. 789–798, Jan. 2016.

\bibitem{Shao2019}
X.~Shao, G.~Tang, and B.-K. Bao, ``Personalized travel recommendation based on sentiment-aware multimodal topic model,'' \emph{IEEE Access}, vol.~7, p. 113043–113052, 2019.

\bibitem{Bin2019}
C.~Bin, T.~Gu, Y.~Sun, and L.~Chang, ``A personalized poi route recommendation system based on heterogeneous tourism data and sequential pattern mining,'' \emph{Multimedia Tools and Applications}, vol.~78, no.~24, p. 35135–35156, Aug. 2019.

\bibitem{Nikookar2022}
S.~Nikookar, P.~Sakharkar, B.~Smagh, S.~Amer-Yahia, and S.~B. Roy, ``Guided task planning under complex constraints,'' in \emph{2022 IEEE 38th International Conference on Data Engineering (ICDE)}.\hskip 1em plus 0.5em minus 0.4em\relax IEEE, 2022.

\bibitem{Alenezi2022}
T.~Alenezi and S.~Hirtle, ``Normalized attraction travel personality representation for improving travel recommender systems,'' \emph{IEEE Access}, vol.~10, p. 56493–56503, 2022.

\bibitem{Chiang2015}
H.-S. Chiang and T.-C. Huang, ``User-adapted travel planning system for personalized schedule recommendation,'' \emph{Information Fusion}, vol.~21, p. 3–17, Jan. 2015.

\bibitem{Benouaret2017}
I.~Benouaret and D.~Lenne, ``Recommending diverse and personalized travel packages,'' in \emph{Lecture Notes in Computer Science}.\hskip 1em plus 0.5em minus 0.4em\relax Springer International Publishing, 2017, pp. 325--339.

\bibitem{Gehlot2022}
A.~Gehlot and R.~Singh, ``Research guide for ml based smart tourist system,'' in \emph{2022 International Interdisciplinary Humanitarian Conference for Sustainability (IIHC)}, 2022, pp. 1427--1434.

\bibitem{Huang2020}
T.~Huang, Y.-J. Gong, Y.-H. Zhang, Z.-H. Zhan, and J.~Zhang, ``Automatic planning of multiple itineraries: A niching genetic evolution approach,'' \emph{IEEE Transactions on Intelligent Transportation Systems}, vol.~21, no.~10, p. 4225–4240, Oct. 2020.

\bibitem{10.1145/3269206.3271739}
H.~Wang, F.~Zhang, J.~Wang, M.~Zhao, W.~Li, X.~Xie, and M.~Guo, ``Ripplenet: Propagating user preferences on the knowledge graph for recommender systems,'' in \emph{Proceedings of the 27th ACM International Conference on Information and Knowledge Management}, ser. CIKM '18.\hskip 1em plus 0.5em minus 0.4em\relax Association for Computing Machinery, 2018, p. 417–426.

\bibitem{jannach2022session}
D.~Jannach, M.~Quadrana, and P.~Cremonesi, ``Session-based recommender systems,'' in \emph{Recommender Systems Handbook}.\hskip 1em plus 0.5em minus 0.4em\relax Springer, 2022, pp. 301--334.

\bibitem{10.5555/501065}
\emph{Software product lines: practices and patterns}.\hskip 1em plus 0.5em minus 0.4em\relax Addison-Wesley Longman Publishing Co., Inc., 2001.

\bibitem{van2010software}
F.~van~der Linden, K.~Schmid, and E.~Rommes, \emph{Software Product Lines in Action: The Best Industrial Practice in Product Line Engineering}.\hskip 1em plus 0.5em minus 0.4em\relax Springer Berlin Heidelberg, 2010.

\bibitem{thum2014featureide}
T.~Th{\"u}m, C.~K{\"a}stner, F.~Benduhn, J.~Meinicke, G.~Saake, and T.~Leich, ``Featureide: An extensible framework for feature-oriented software development,'' \emph{Science of Computer Programming}, vol.~79, pp. 70--85, 2014.

\bibitem{d2023democratizing}
G.~d'Aloisio, A.~Di~Marco, and G.~Stilo, ``Democratizing quality-based machine learning development through extended feature models,'' in \emph{International Conference on Fundamental Approaches to Software Engineering}.\hskip 1em plus 0.5em minus 0.4em\relax Springer, 2023, pp. 88--110.

\bibitem{forsyth_top_2021}
A.~Forsyth, ``\BIBforeignlanguage{en}{Top 5 {Benefits} of {Low}-{Code}},'' Jan. 2021.

\bibitem{WONG2023253}
I.~A. Wong, Q.~L. Lian, and D.~Sun, ``Autonomous travel decision-making: An early glimpse into chatgpt and generative ai,'' \emph{Journal of Hospitality and Tourism Management}, vol.~56, pp. 253--263, 2023.

\bibitem{BENADDI2024275}
L.~Benaddi, C.~Ouaddi, A.~Jakimi, and B.~Ouchao, ``Towards a software factory for developing the chatbots in smart tourism mobile applications,'' \emph{Procedia Computer Science}, vol. 231, pp. 275--280, 2024, (EUSPN/ICTH 2023).

\bibitem{wohlin_experimentation_2012}
C.~Wohlin, P.~Runeson, M.~H\"{o}st, M.~C. Ohlsson, B.~Regnell, and A.~Wessl\'{e}n, \emph{\BIBforeignlanguage{en}{Experimentation in {Software} {Engineering}}}.\hskip 1em plus 0.5em minus 0.4em\relax Springer, 2012.

\end{thebibliography}
